\documentclass[10pt,twocolumn]{IEEEtran}
\usepackage{multicol}

\usepackage{color}
\usepackage{xcolor}
\usepackage{soul}
\usepackage{amsmath}
\usepackage{graphics}
\usepackage{setspace}
\usepackage{cite}
\usepackage{latexsym}
\usepackage{float}
\usepackage{epsfig}
\usepackage{multirow}
\usepackage{cite,cases,url}
\usepackage{amssymb}
\usepackage{graphicx}
\usepackage{epstopdf}
\usepackage{balance}
\usepackage{subcaption}
\usepackage{enumerate}
\usepackage{array}
\usepackage{algorithmic}
\usepackage{algorithm}
\usepackage{enumitem}
\usepackage{dblfloatfix}
\usepackage[normalem]{ulem}
\useunder{\uline}{\ul}{}

\newcolumntype{L}{>{\centering\arraybackslash}m{5cm}}
\newcolumntype{K}{>{\centering\arraybackslash}m{6cm}}
\newcolumntype{P}{>{\centering\arraybackslash}m{2.3cm}}
\newcolumntype{M}{>{\raggedright\arraybackslash}m{2cm}}
\newcolumntype{N}{>{\raggedright\arraybackslash}m{2.5cm}}

\usepackage{tikz}
\newcommand*\circled[1]{\tikz[baseline=(char.base)]{
            \node[shape=circle,draw,inner sep=1pt] (char) {#1};}}
\begin{document}
\bstctlcite{bstctl:nodash}

\title{
End-to-End O-RAN Security Architecture, Threat Surface, Coverage, and the Case of the Open Fronthaul}


\author{\IEEEauthorblockN{Aly Sabri Abdalla and Vuk Marojevic} \\
\IEEEauthorblockA{Dept. Electrical and Computer Engineering, Mississippi State University,
Mississippi State, MS, USA\\
asa298@msstate.edu, vuk.marojevic@msstate.edu}

\vspace{-8mm}
}

\maketitle

\begin{abstract}

O-RAN establishes an advanced radio access network (RAN) architecture that supports inter-operable, multi-vendor, and artificial intelligence (AI) controlled wireless access networks. 
The unique components, interfaces, and technologies of O-RAN 
differentiate it from the 
3GPP RAN. 
Because O-RAN supports 3GPP protocols, currently 4G and 5G, while offering additional network interfaces and controllers, it has a larger attack surface. 
The O-RAN security requirements, vulnerabilities, threats, and countermeasures must be carefully assessed for it to become a platform for 5G Advanced and future 6G wireless. 
This article presents the ongoing standardization activities of the O-RAN Alliance for modeling the potential threats to the network and to the open fronthaul interface, in particular. 
We identify end-to-end 
security threats and discuss those on the open fronthaul in more detail. We then provide recommendations for countermeasures to tackle the identified security risks and 
encourage industry to establish 
standards and best practices for safe
and secure implementations of the open fronthaul interface. 

\end{abstract}

\IEEEpeerreviewmaketitle
\section{Introduction}
\label{sec:intro}
Recent advances in wireless communications are expected to enable a fully mobile and connected society in the 
future, which will be characterized by the tremendous growth in connectivity, traffic volume, and a much broader range of services. 
Besides the market requirements, the mobile communication society also requires a sustainable development of the ecosystem which produces the need for further improved system efficiencies, including spectrum efficiency, energy efficiency, operational efficiency, and cost efficiency. 


As the preparation of the fourth release of 5G---5G Advanced---
is underway, the companies and partner organizations identify various technical topics that need to be researched in 
Release 18 \textcolor{black}{and future releases}. 
Unlike Releases 16 and 17, which helped extend 5G to new verticals, the objective of 
Release 18 is covering more demanding applications, such as truly mobile extended reality services, 
network intelligence, and enhanced support for new use cases. Therefore, new technologies and paradigm shifts are expected for supporting the critical aspects of next-generation wireless communication networks as part of 5G Advanced and 6G standardization efforts. 

The current 5G architecture has been designed 
on the basis of network virtualization, Cloud technology, and software-defined networks (SDNs) to enable agile and self-adapting network solutions. 
5G considers edge computing and other critical architectural refinements over previous network generations~\cite{niknam2020intelligent}; however, the deployment follows the trend of previous generations and may lead to 5G becoming a by and large monolithic and inflexible infrastructure with vendor lock-in. 
\textcolor{black}{Next-generation network innovation requires the} transformation to a flexible, agile and disaggregated architecture to support service heterogeneity, coordination among multiple technologies, and rapid on-demand deployments. 
Such transformations are enabled by the emerging open radio access network (O-RAN). 

O-RAN is an industry-driven architecture \textcolor{black}{that is maintained} by the O-RAN Alliance. \textit{Openness} and \textit{intelligence} are among the main characteristics of the O-RAN architecture that enables multi-vendor, inter-operable, artificial intelligence (AI) empowered hierarchical networks
~\cite{Pratheek}. Network disaggregation is one of the O-RAN \textcolor{black}{architectural} features that has been elevated by decoupling network functions and harnessing SDN/NFV principles~\cite{SDR}. The unique components, interfaces, and technologies of O-RAN differentiate it from those of 3GPP, while leveraging the 3GPP framework and protocols. The security challenges and resultant risks of O-RAN will \textcolor{black}{therefore} be different \textcolor{black}{from} those of legacy 3GPP networks and  
must be carefully studied to reduce 
risks,   
\textcolor{black}{vulnerabilities, and exposures to attacks and misconfigurations}. 

The O-RAN Alliance defines study items that are currently organized into \textcolor{black}{eleven} technical workgroups (WGs) and four focus groups (FGs). One of the WGs is the security WG (SWG) which is committed to developing the security requirements, designs, and solutions that enable an open, interoperable, and secure O-RAN system. The SWG established itself in the second quarter of 2021 and 
has since developed initial work items specifying initial O-RAN security threat models and security protocols to be addressed when building a secure end-to-end O-RAN system~\cite{SFGThreats, SFGReqs}. Academics have started to evaluate the security of O-RAN. The work presented in~\cite{ORANSEC1} has demonstrated a security evaluation 
of the O-RAN architecture with a threat analysis for different domains. Reference~\cite{ORANSEC2} shows a security vulnerability case study revealing the missing authentication and authorization of O-RAN interfaces and components.

This paper compiles the threat vectors against the major O-RAN components: the open fronthaul, the near and non-real time RAN intelligent controllers (RICs), the service management and \textcolor{black}{orchestration} 
(SMO), the O-RAN Cloud, and the machine learning (ML)/AI \textcolor{black}{employed by xApps and rApps}. We then focus our detailed threat analysis on O-RAN's open fronthaul interface and services, and \textcolor{black}{introduce} the 
security countermeasures 
to tackle the identified security risks. 
Our initial research provides recommendations and guidelines for consideration and further evaluation, leading to 
standards and best practices for safe
and secure O-RAN implementation and operation. 

The rest of the paper is organized as follows: Section II 
\textcolor{black}{presents} the O-RAN architecture, including its components and interfaces, with focus on security threats and principles. Section III discusses the O-RAN open fronthaul threats. 
Section IV introduces the 
security countermeasures 
to preserve the security requirements of the open fronthaul given the previously identified threats. 
Section V provides the concluding remarks.

\section{
End-to-End O-RAN Security Architecture, Threats, and \textcolor{black}{Security Coverage} 
} 
\label{sec:ORANARCH}
\begin{figure}[t]
    \centering
    \includegraphics[height=9.2 cm, width=0.48\textwidth]{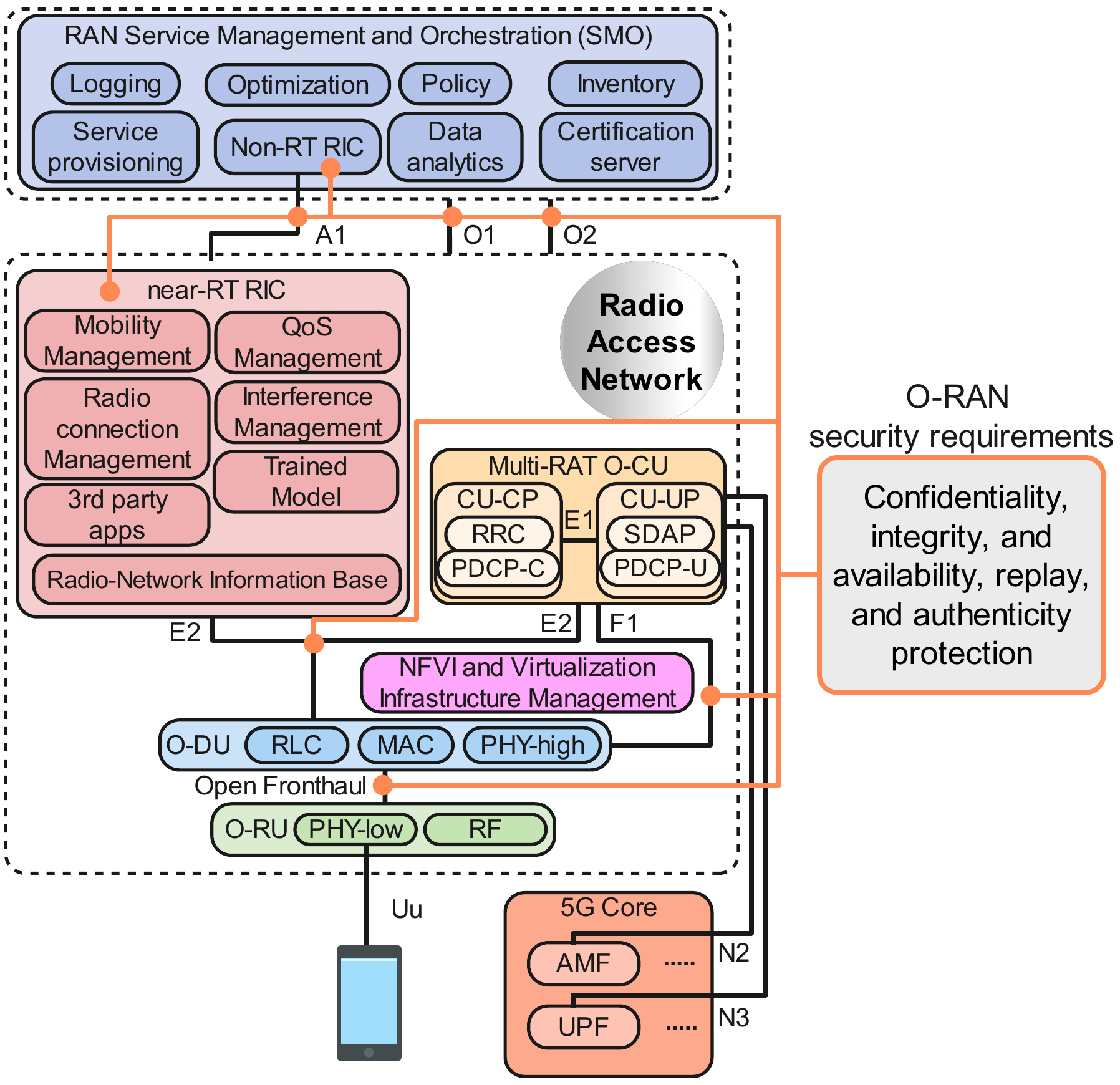}
    \caption{
    The O-RAN architecture with security requirements.}
    \label{fig:Figure1}
    \vspace{-3mm}
\end{figure}
\textcolor{black}{This section introduces the O-RAN security 
requirements. 
Fig. 1 presents the O-RAN architecture and identifies the general security needs.}
The O-RAN architecture is divided into two main subsystems. These are the 
RAN and the 
SMO
. 
The logical components of the RAN subsystem include 
the O-RAN \textcolor{black}{central} unit (O-CU), the O-RAN distributed unit (O-DU), and the O-RAN radio unit (O-RU). 
In addition, there are two RAN Intelligent Controllers (RICs) 
that cover different timescales: 
the near-real time RIC (near-RT RIC), which is part of the RAN \textcolor{black}{subsystem}, and the non-RT RIC, which is part of the SMO. 
\textcolor{black}{O-RAN micro-services called xApps and rApps implement near-RT and non-RT \textcolor{black}{control} services \textcolor{black}{as part of the near-RT and non-RT RICs}, respectively \cite{RezaGC22}. }

O-RAN defines open interfaces to handle the data and control flows between the O-RAN components, which may be implemented as virtual machines and execute on a single compute nodes or on networked nodes. \textcolor{black}{The open fronthaul 
establishes the 
\textcolor{black}{interface} between the O-RU and the O-DU~\cite{ORANWG4}.} The F1 interface 
connects the O-CU to the O-DUs. The E1 interface enable coordination between the O-CU control and user planes. The 
E2 interface forwards the measurements from the O-DU and O-CU to the near-RT RIC and the configuration commands 
to the O-CU and O-DU. 
\textcolor{black}{User or RAN-specific data 
can be fed to the xApps 
for data processing employing AI}.
The O1 interface is accountable for collecting the data from all of the connections for operations \& administration functions. 
\textcolor{black}{Policy guidance can be transmitted via the A1 interface to xApps implementing AI models that may support RAN applications} such as network slicing, quality of service (QoS) and resource management, and mobility management~\cite{ORANAly}.

End-to-end security is a mandatory feature and must be maintained and managed across all components and interfaces of the O-RAN architecture. However, the openness, disaggregation of network functionalities, and intelligence are the unique properties of O-RAN and facilitate the integration of 
new functions, protocols, components, and interfaces. This openness expands the threat surface 
and makes O-RAN prone to additional security risks beyond those of the 3GPP architecture. 
\textcolor{black}{Table~\ref{tab:Table1} presents the major O-RAN security risks and threats 
that have been 
\textcolor{black}{identified} by the SWG~\cite{SFGThreats, SFGReqs}.} 
Table~\ref{tab:Table1} \textcolor{black}{also identifies the affected components and} 
the main security principles that should be taken into consideration while defining security requirements, recommendations, and countermeasures 
in future standardization efforts.     

With the expanded threat surface of O-RAN, different security features across
the O-RAN architecture \textcolor{black}{need} to be identified to protect the critical assets and maintain the integrity, availability, confidentiality, replay protection, and authenticity. In continuation, we introduce the security features as defined for the main O-RAN components~\cite{SFGThreats, SFGReqs}. 
\begin{table*}
\vspace{-40 pt}
\centering
\caption{O-RAN threats and coverage security principles\textcolor{black}{~\cite{SFGThreats, SFGReqs}}.}
\footnotesize
\label{tab:Table1}
\centering
{\begin{tabular}{|p{0.8cm}|p{1.0cm}|p{6.8cm}|p{2.5cm}|p{4.3cm}|}
\hline
\multicolumn{2}{|c|}{\textbf{O-RAN Component}}  &\textbf{Threat } &  \textbf{Affected Components} &  \textbf{Security Principles} 
\\ \hline
\textbf{Front-haul}
& 
Fronthaul 
&
\textcolor{black}{- {An attacker penetrates O-DU and beyond through O-RU 
}} 
\newline - {Unauthorized access to the open fronthaul Ethernet L1 physical layer interface}
& 
rApps, xApps, O-RU, O-DU, O-CU, Near-RT RIC, Non-RT RIC
&
Mutual authentication, access control, cryptography, key management, public key infrastructure (PKI), trusted communication, and security assurance 
\\
\cline{2-4}
&
M-Plane 
&
\textcolor{black}{- {An attacker attempts to intercept the fronthaul (man in the middle---MITM---attack) over the M-Plane}} 
& 
rApps, xApps, O-RU, O-DU, O-CU, Near-RT RIC, Non-RT RIC
&
\\

\cline{2-4}
& 
{S-Plane} 
& 
\textcolor{black}{- {Denial of service (DoS) attack against a master clock }} 
\newline - {Impersonation of a master clock (spoofing) within a precision time protocol (PTP) network with a fake ANNOUNCE message}
\newline - {Rogue PTP instance 
\textcolor{black}{claiming the role of a grand master clock}}
\newline - {Selective interception and removal of PTP timing packets}
\newline - {Packet delay manipulation attack}
&
O-DU, O-RU
&
\\
\cline{2-4}

& 
C-Plane  
& 
\textcolor{black}{- {Spoofing of downlink (DL) C-plane messages}}
\textcolor{black}{- {Spoofing of uplink (UL) C-plane messages}}
& 
O-DU, O-RU
&
\\
\cline{2-4}

& 
U-Plane  
& 
\textcolor{black}{- {An attacker attempts to intercept the fronthaul (MITM) over U-plane}} 
& 
O-DU, O-RU
&
\\
\hline
\textbf{Near-RT RIC} 
& 
Near-RT RIC 
& 
\textcolor{black}{- Malicious apps can exploit UE identification, track UE location and change UE priority}
&
Near-RT RIC, UE, xApps
&
Secure boot and self-configuration, security management of open source software, privacy assurance, continuous logging, monitoring and vulnerability handling, robust isolation, secure Cloud computing and virtualization
\\

\cline{2-5}
& \rule{0pt}{4ex}    
{xApp}
& 
\textcolor{black}{- An attacker exploits xApps vulnerabilities and misconfiguration} 
\newline \textcolor{black}{- Conflicting xApps unintentionally or maliciously impact O-RAN system functions to degrade performance or trigger DoS}
\newline - An attacker compromises xApp isolation
&
O-CU, Near-RT RIC, xApps
&
Mutual authentication, access control, cryptography, key management, PKI, robust isolation, secure Cloud computing and virtualization, recoverability \& backup
\\
\hline
\textbf{Non-RT RIC} 
& 
Non-RT RIC 
& 
\textcolor{black}{- {An attacker penetrates the non-RT RIC to cause DoS or degrade the performance} }
\newline \textcolor{black}{- UE sniffing in the Non-RT RIC}
&
Non-RT RIC, rApps, UE
&
Mutual authentication, access control, cryptography, key management, PKI, privacy assurance, security management of open source software, secure Cloud computing and virtualization, secure boot and self-configuration
\\

\cline{2-5}

& 
rApp & 
\textcolor{black}{- {An attacker exploits rApps vulnerabilities and misconfiguration} } 
\newline \textcolor{black}{- An attacker bypasses authentication and authorization}
\newline \textcolor{black}{- An attacker compromises rApp isolation}
\newline \textcolor{black}{- Conflicting rApps unintentionally or maliciously impact O-RAN system functions to degrade performance or trigger DoS}
&
rApps, UE, Non-RT RIC, Near-RT RIC, xApps
&
Mutual authentication, access control, cryptography, key management, PKI, privacy assurance, secure boot and self-configuration, robust isolation, secure Cloud computing and virtualization, continuous logging, monitoring and vulnerability handling. 
\\
\hline
\multicolumn{2}{|c|}{\textbf{SMO}} 
&  
\textcolor{black}{- {An attacker can exploit the improper/missing authentication 
of SMO functions} } 
\newline \textcolor{black}{- Overload DoS attack 
}
&
SMO components
&
Mutual authentication, access control, privacy assurance, recoverability \& backup, privacy assurance, continuous logging, monitoring and vulnerability handling, secure update 
\\
\hline
\multicolumn{2}{|c|}{\textbf{O-Cloud}} 
&
{
\textcolor{black}{- An attacker compromises VNF/CNF images and embedded secrets }}
\newline \textcolor{black}{- An attacker exploits weak orchestrator configuration, access control and isolation}
\newline \textcolor{black}{- Misuse of a virtual machine (VM) or container (CN) to attack other VM/CN, hypervisor/container engine, other hosts (memory, network, storage), etc.}
\newline \textcolor{black}{- Spoofing of and eavesdropping on network traffic}
\newline \textcolor{black}{- An attacker compromises auxiliary/supporting network and security services}
& 
O-Cloud components
&
Mutual authentication, access control, privacy assurance, recoverability \& backup, privacy assurance, continuous logging, monitoring and vulnerability handling, secure update, secure boot and self-configuration, robust isolation, secure storage 
\\
\hline
\multicolumn{2}{|c|}{\textbf{ML/AI}} 
&
{
\textcolor{black}{- Poisoning the ML training data (data poisoning attack) }}
\newline \textcolor{black}{- An attacker exploits weak orchestrator configuration, access control and isolation}
\newline \textcolor{black}{- ML model alteration (system manipulation and compromise ML data confidentiality and privacy)}
\newline \textcolor{black}{- Spoofing of and eavesdropping on network traffic}
\newline \textcolor{black}{- Transfer learning attack}
& 
Near-RT RIC, Non-RT RIC, xApps, rApps
&
Mutual authentication, access control, continuous logging, monitoring and vulnerability handling, recoverability \& backup, privacy assurance, secure boot and self-configuration, secure update, secure Cloud computing and virtualization
\\
\hline
\end{tabular}
\vspace{-3 mm}
}
\end{table*}
\begin{itemize}
    \item \textbf{Fronthaul interface domain security:} A set of features and mechanisms that enable a secure flow of critical control, user, management, and synchronization plane data, such as timing configuration, troubleshooting and trace logs, and user data, that are transported over the fronthaul, interconnecting multiple O-RUs and O-DUs.
    \item \textbf{Near-RT RIC domain security:} A set of features and mechanisms that safeguard data transported \textcolor{black}{toward} the near-RT RIC which then optimizes the overall RAN performance over the following interfaces: \circled{1} the policies applicable to UEs and cells and A1 enrichment information delivered over the A1 interface and \circled{2} the persistent configuration used by the near-RT RIC to control the RAN, xApp-related messages, near-RT RIC services messages and policies used 
    to monitor, suspend/stop, override or control the behavior of the E2 node delivered over the O1 and E2 interfaces. 
    \item \textbf{Non-RT RIC domain security:} A set of features and mechanisms that ensure \circled{1} secure creation, modification, and management of A1 policies and A1 enrichment information that is collected or derived at the SMO/non-RT RIC and \circled{2} authentication and authorization of the discovery and request of A1 enrichment information from the near-RT RIC.
    \item \textbf{xApp and rApp domain security:} A set of features and mechanisms for protecting \circled{1} the training or test data, \circled{2} the trained ML model, and \circled{3} the expected  prediction outcomes. The data sets may be collected externally or internally to the near-RT RIC, O-CU, O-DU\textcolor{black}{, and O-RU} and passed to the ML training hosts to be applied \textcolor{black}{by an xApp or rApp}. 
    The trained ML model may include intellectual property \textcolor{black}{information}, numerous 
    hyperparameters and millions of learned parameters. 
    The expected prediction outcomes of the ML model and the behavior of the ML system \textcolor{black}{include} tasks for data collection, data wrangling, pipeline management, model retraining, and model deployment. 
    \item \textbf{O-RU domain security:} A set of features and mechanisms that facilitate the secure exchange of reference signals, synchronization signals, and data and control channels in the downlink and uplink between the O-RUs and UEs.
     \item \textbf{O-CU and O-DU domain security:} A set of features and mechanisms that maintain the integrity of the 3GPP application-related data such as subscription data, session data, call control and inter and intra-slice UE priority related information. 
     \item \textbf{O-Cloud domain security:} A set of features and mechanisms that need to be provisioned to minimize risk exposure affecting 
     telemetry information of O-Cloud deployments, O-Cloud provisioning information, and O-Cloud software management information.  
\end{itemize}

\section{
\textcolor{black}{Open Fronthaul} Security Threats} 
\label{sec:contribution}
In this section we examine the potential threats targeting the 
open fronthaul interface and the management, control, user, and synchronization data planes transmitted over the open fronthual interface within the O-RAN architecture and provide 
\textcolor{black}{a} risk analysis associated with each threat.  
\subsection{Threats against the fronthaul interface}
The first threat category 
is related to 
attack exploit vulnerabilities of the open fronthaul interface. 
\begin{figure}[t]
    \centering
    \includegraphics[ width=0.49\textwidth]{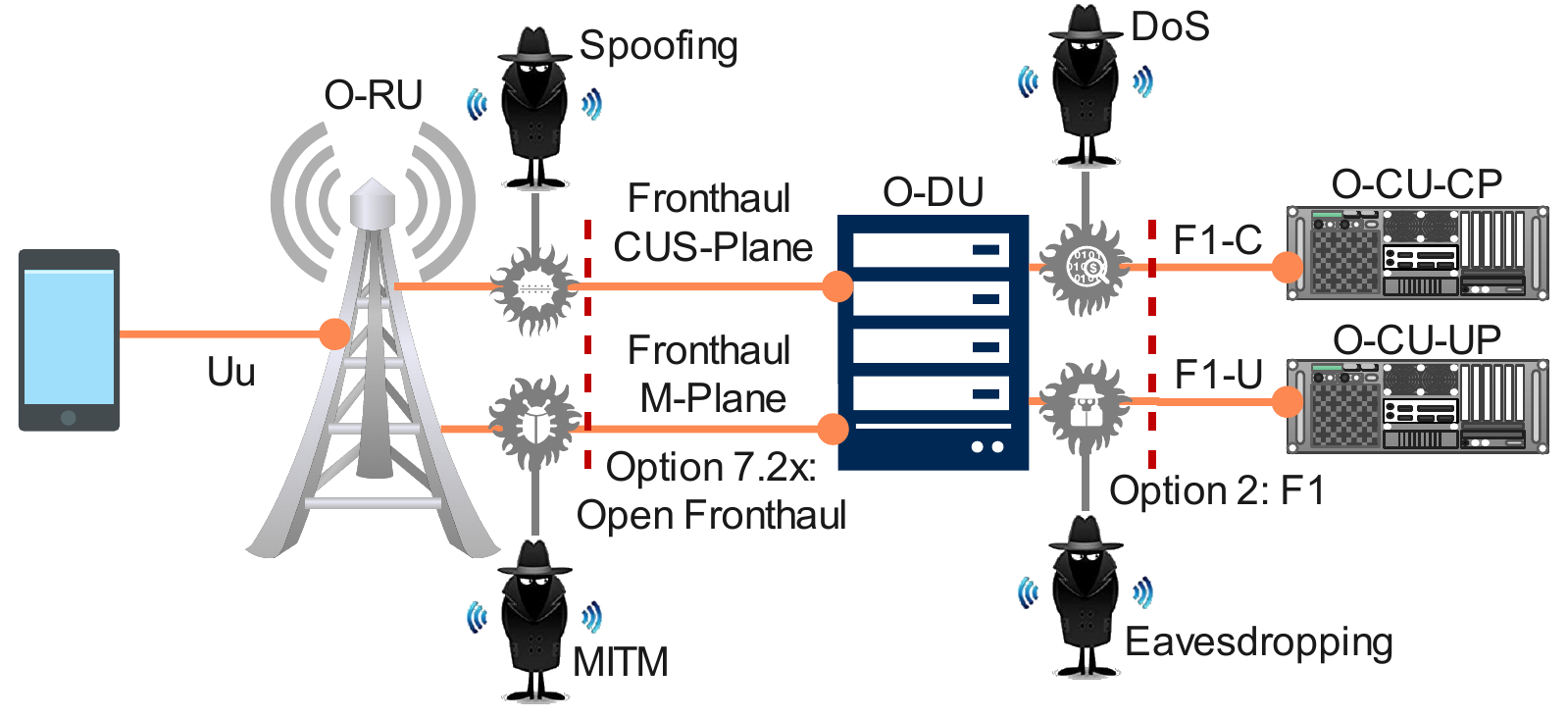}
    \caption{
    The O-RAN fronthaul.}
    \label{fig:FigureFH}
    \vspace{-3mm}
\end{figure}
Functional splits were introduced in 3GPP Release~15 to allow splitting the base station functionalities into the CU and the DU, implementing the higher and lower layers of the RAN protocol stack, respectively. Although 3GPP defines many split options, {vendors} use proprietary implementations and interfaces which has led to single vendor network solutions. O-RAN adopts functional split Option 2 for the F1 interface between the O-DU and O-CU, and Option 7.2x for implementing the fronthaul, the interface between the O-DU and O-RU as shown in~Fig.~\ref{fig:FigureFH}.

\textcolor{black}{O-RAN employs a modified version of the common public radio interface (CPRI), the enhanced CPRI (eCPRI), for the fronthaul~\cite{ORANFH}. The eCPRI interface allows for the separation of the radio unit and the baseband unit, and enables the use of off-the-shelf networking equipment
, which can reduce the costs and increase the scalability of the network. 
} This 
facilitates having different vendors for the O-RUs and the O-DUs,  
\textcolor{black}{which need} to be managed as different entities that may have heterogeneous security levels. 
The O-DU will have to bridge the 
control and other data traffic 
between the management 
O-CU 
and the O-RU as shown in Fig.~\ref{fig:Figure1}. Hence the possibilities to reach and penetrate the northbound systems (O-CU, near-RT RIC, and SMO) beyond the O-DU through the open fronthaul interface becomes a possible vulnerability in this split architecture that can be exploited by an attacker. 

Another 
\textcolor{black}{threat is the unauthorized} access to \textcolor{black}{the physical layer, or layer 1 (L1), of the open fronthaul} 
by comprising one or more coaxial cables, twisted pairs, RF links, or optical fibers. Each end of the physical 
interface encompasses a physical connection (an Ethernet port) to physical O-RAN network elements, the O-DU and the O-RU. An attacker who gains 
physical access to the 
fronthaul interface 
can launch attacks that can compromise the availability, integrity, and confidentiality of the 
interface. 
Specifically, an unauthorized device on the fronthaul 
interface can 
\begin{enumerate}
    \item Flood the L1 interface with 
    network traffic causing disruption or degradation of authorized network elements on the fronthaul interface,
    \item Transmit L2 packets to authorized network devices causing disruption 
    or degradation of the fronthaul interface performance,
    \item Deny services by disabling a physical connection to a network element either by removing an Ethernet port connection or cutting the physical interface, and
    \item Access \textcolor{black}{and manipulate} the management, synchronization, control, and user plane traffic. 
\end{enumerate}

\subsection{Threats against the management, synchronization, control, and user planes}
According to the O-RAN fronthaul specifications, there are four types of planes 
that support the various functionalities of the O-RAN fronthaul: management plane (M-plane), synchronization plane (S-plane), user plane (U-plane), and control plane (C-plane). 

\circled{I} \textcolor{black}{The M-plane 
provides a variety of 
management functions to set O-RU parameters 
as required by the C/U-plane and S-plane, manage the O-RU software, perform fault management, and so forth.} The O-RAN fronthaul specifications for the M-plane provide various parameters 
for fault, configuration, accounting, performance, and security (FCAPS) functions. The 7.2x functional split 
\textcolor{black}{requires} high bit rate transmissions with strict bandwidth, latency, and 
transport link 
performance requirements. \textcolor{black}{This 
limits the options for employing extensive security measures} within the O-RAN system because of the \sout{\textcolor{black}{the}} 
processing delay that they would incur. 
As a result, the O-RAN security risks increase, specifically for the 
M-plane, where an attacker may be capable of launching man-in-the-middle (MITM), passive wiretapping, or denial of service (DoS) attacks over the fronthaul 
interface to intercept the M-plane messaging after gaining unauthorized access to the operations and maintenance (OAM).

\circled{II} The S-plane is responsible for the timing and synchronization of messages between the O-DU and O-RU. Through the S-plane, the O-RAN fronthaul specifications support protocols such as precision time protocol (PTP) and synchronous Ethernet (SyncE) to achieve high-accuracy synchronization between the O-RU and the O-DU, which supplies the 
the master clock. 
The S-plane must be protected against DoS attacks targeting the master clock of the timing network used by the fronthaul to maintain availability and  accuracy of the O-RAN system. An attacker can attack the master clock by sending an excessive number of time protocol packets. 
Such an attack may result in a situation where the clock service is interrupted 
or the timing protocol is operational but slaves are being provided inaccurate timing information due to the degraded performance of the master clock. 
An attacker within the PTP network \textcolor{black}{may also impersonate} the 
grand-master (GM) clock’s identity and announce itself as a GM candidate by either modifying the in-flight protocol packets or by injecting fake ANNOUNCE messages. 

\textcolor{black}{A GM impersonation attack can result in 
the PTP remaining operational and all clocks synchronized, but 
inaccurate timing information being intentionally distributed.} The attacker may be residing either within the attacked network as an insider or on an external network connected to the network under attack. 
An attacker can also position itself in such way 
to allow intercepting and removing valid synchronization packets. This could lead to clock synchronization errors of all clocks downstream or trigger the free-running mode. Attacks may be launched close to the GM by tapping the egress traffic of an active GM clock to effect a larger set of slaves who depend on this GM for \textcolor{black}{time} synchronization. Attacks may also target one or more slaves by tapping the ingress traffic of a particular slave. Alternatively, an MITM attack may be launched from 
an intermediate node such as a transparent clock (TC), router, or switch. 
\textcolor{black}{This implies that the attacker has obtained} physical access to a node in the PTP network or has gained full control of one device in the network. 
Selective interception and removal can impact timing packets and cause clock degradation of the attacked nodes. Removing all packets or random packets may trigger 
the free running mode~\cite{SFGThreats, SFGReqs}. 

Another threat on the S-plane is the packet delay manipulation 
that may disrupt the symmetric delays between the GM \textcolor{black}{clock} and the slaves. 
An attacker would launch this attack by either tapping the transmission network or by taking control of intermediate nodes such as routers, switches, or TCs 
\textcolor{black}{and provide intentionally inaccurate timing information to slaves. Clock service disruption or time accuracy degradation would occur} and may cause DoS to applications 
that rely on accurate time or potentially bringing down an entire cell. A cell outage caused by misaligned timing may further impact performance of 
neighboring cells.

\circled{III} The U-plane is responsible for the efficient \textcolor{black}{user} data transfer of \textcolor{black}{in-phase and quadrature (IQ)} samples within the strict time limits of the 5G numerology. For the transported U-plane data, an attacker 
may launch 
eavesdropping and DoS attacks 
after breaking the \textcolor{black}{packet data convergence protocol (PDCP)} security prior to any content access. 3GPP defines U-plane integrity protection algorithms in their specifications
, but many of the OEMs may not implement them because they are resource demanding and may impact the user experience, specifically the download and upload data rates. Enabling U-plane integrity protection requires considerable computing resources and adds overhead that directly impacts the maximum throughput that can be measured on the user device. The integrity protection is enabled on the C-plane messages but that still leaves the user’s data traffic vulnerable because the C-plane and U-plane are segregated. For example, the lack of uplink integrity could enable a rogue base station to manipulate the user data messages 
and redirect a user to a malicious website.

\circled{IV} 
C-plane messages
pass from the O-DU to O-RUs and carry information about the scheduling, the coordination required for data transfer, beamforming, mixed numerology, and PRACH handling parameters to be employed when transmitting and receiving IQ sample sequences included in the U-plane message. 

A downlink C-plane message covering multiple symbols must arrive at the O-RU before the end of the downlink U-plane receive window. 
The lack of authentication could allow an adversary to inject fake downlink C-plane messages that falsely claim to be from the associated O-DU. This could result in blocking the O-RU from processing the corresponding U-plane packets, leading to temporary DoS. 

A uplink C-plane message includes multiple symbols that must be delivered to the O-RU prior to the reception of the earliest air interface uplink \textcolor{black}{U-plane} signal sample.  
Uplink C-plane messages from the O-RU to the O-DU are only defined for 
LTE licensed assisted access~(3GPP TR 36.889) and New Radio-Unlicensed~(3GPP TR 38.889) operations; otherwise, there should be no C-plane messages originating at the O-RU. The lack of authentication 
\textcolor{black}{may result in an adversary injecting} fake uplink C-plane messages \textcolor{black}{into O-RUs}. 
This may lead to 
reduced cell performance or even DoS 
encompassing all O-RUs that are associated with 
\textcolor{black}{the mimicked O-DU}. 


\section{Fronthaul Security Coverage
}
\label{sec:Challenges}
\begin{figure}[t]
    \centering
    \includegraphics[ width=0.49\textwidth]{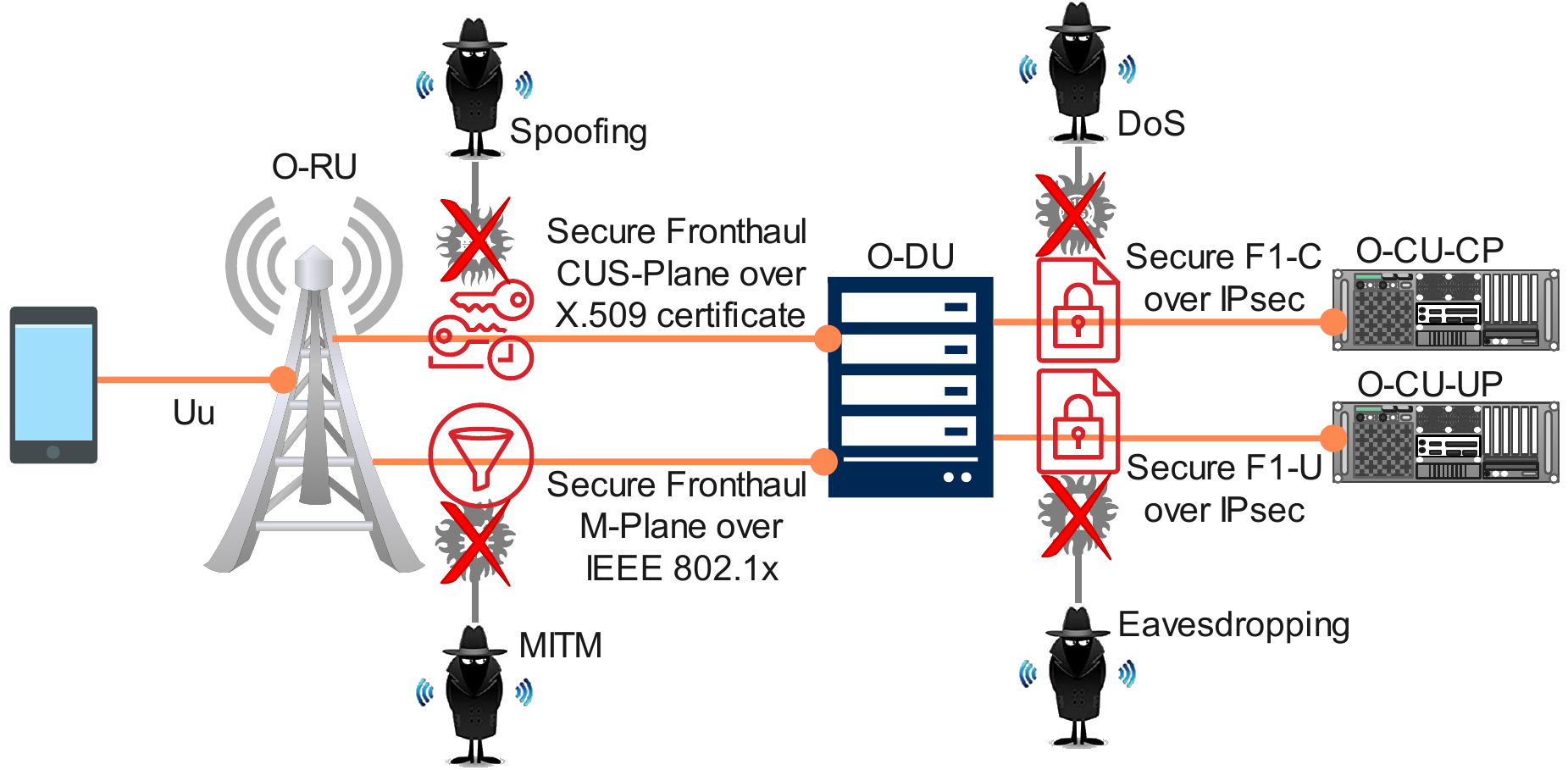}
    \caption{
    Secured O-RAN fronthaul interface.}
    \label{fig:FigureFHSecure}
\end{figure}
This section discusses 
the possible set of security features and mechanisms that are of critical importance to be implemented within the O-RAN system for tackling potential threats to the fronthaul interface and the corresponding 
management, control, user, and synchronization data planes. The provided security solutions in this section are expected to 
\textcolor{black}{initiate further analysis and 
refinement for devising security requirements, recommendations, and potential countermeasures as part of future O-RAN 
architectural revisions.} 
In what follows, \textcolor{black}{we 
identify} different mutual authentication solutions that need to be established for the O-RAN system of Fig.~\ref{fig:FigureFHSecure} to be able to verify who does what as a means to 
\textcolor{black}{detect 
fake base stations, unauthorized or malicious components, 
applications, 
users and administrators}. Mutual authentication that is capable of filtering unauthorized/unexpected traffic \textcolor{black}{flowing through} O-RAN components and 
interfaces, can restrict access to component configurations and provide legitimate access to the hardware and software to maintain the trust chain. In continuation we elaborate on certificate-based authentication, port-based authentication, and IP security-based authentication.

\begin{figure*}[ht]
    \centering
    \includegraphics[width=0.95\textwidth]{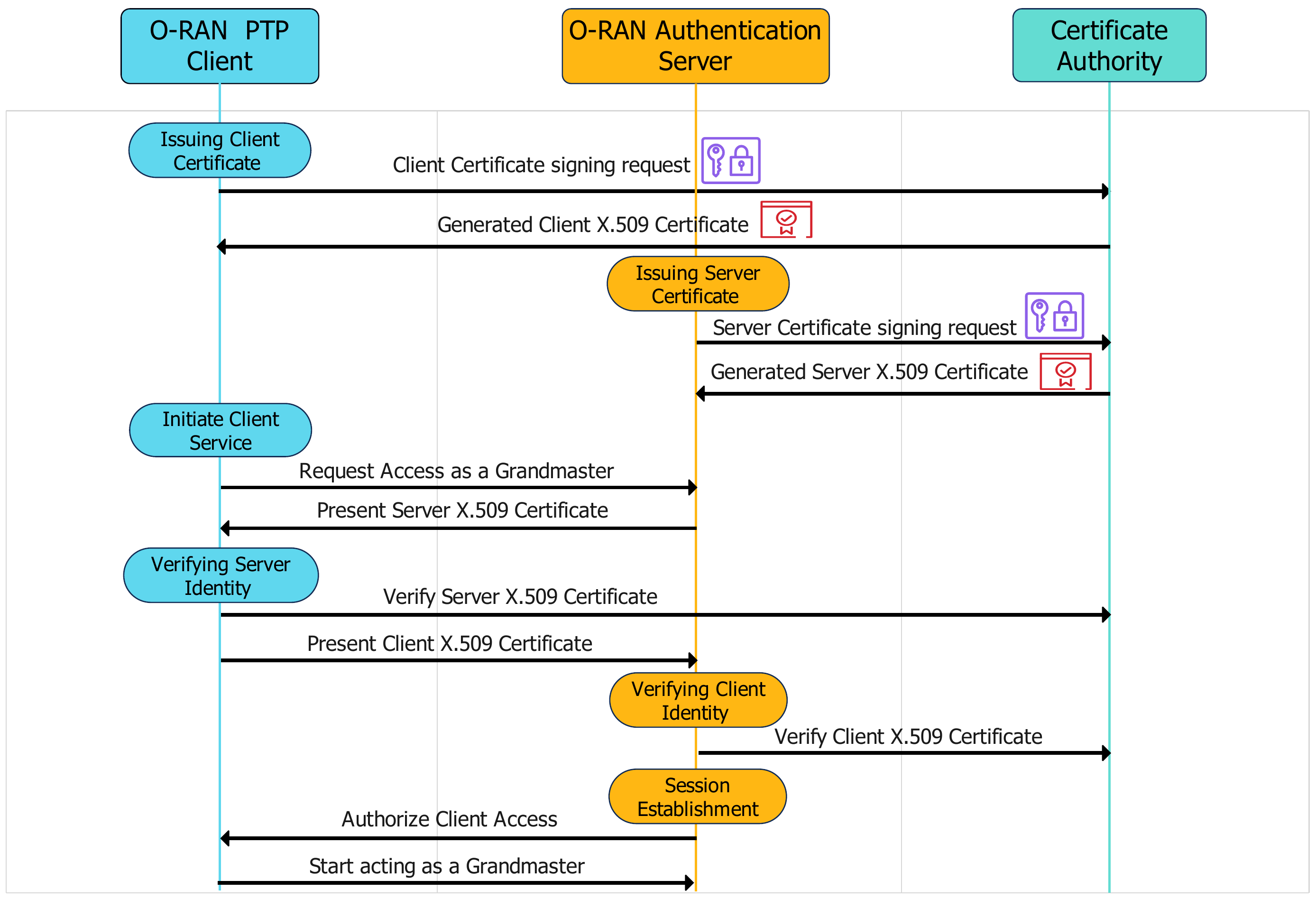}
    \vspace{22mm}
    \label{fig:FigureORANX509}
    \vspace{-14mm}
    \caption{Mutual authentication between \textcolor{black}{the} O-RAN PTP client and \textcolor{black}{the} O-RAN authentication server via X.509 certificates }
\end{figure*}
       \begin{figure*}[h!]
    \centering
    \includegraphics[ width=0.95\textwidth]{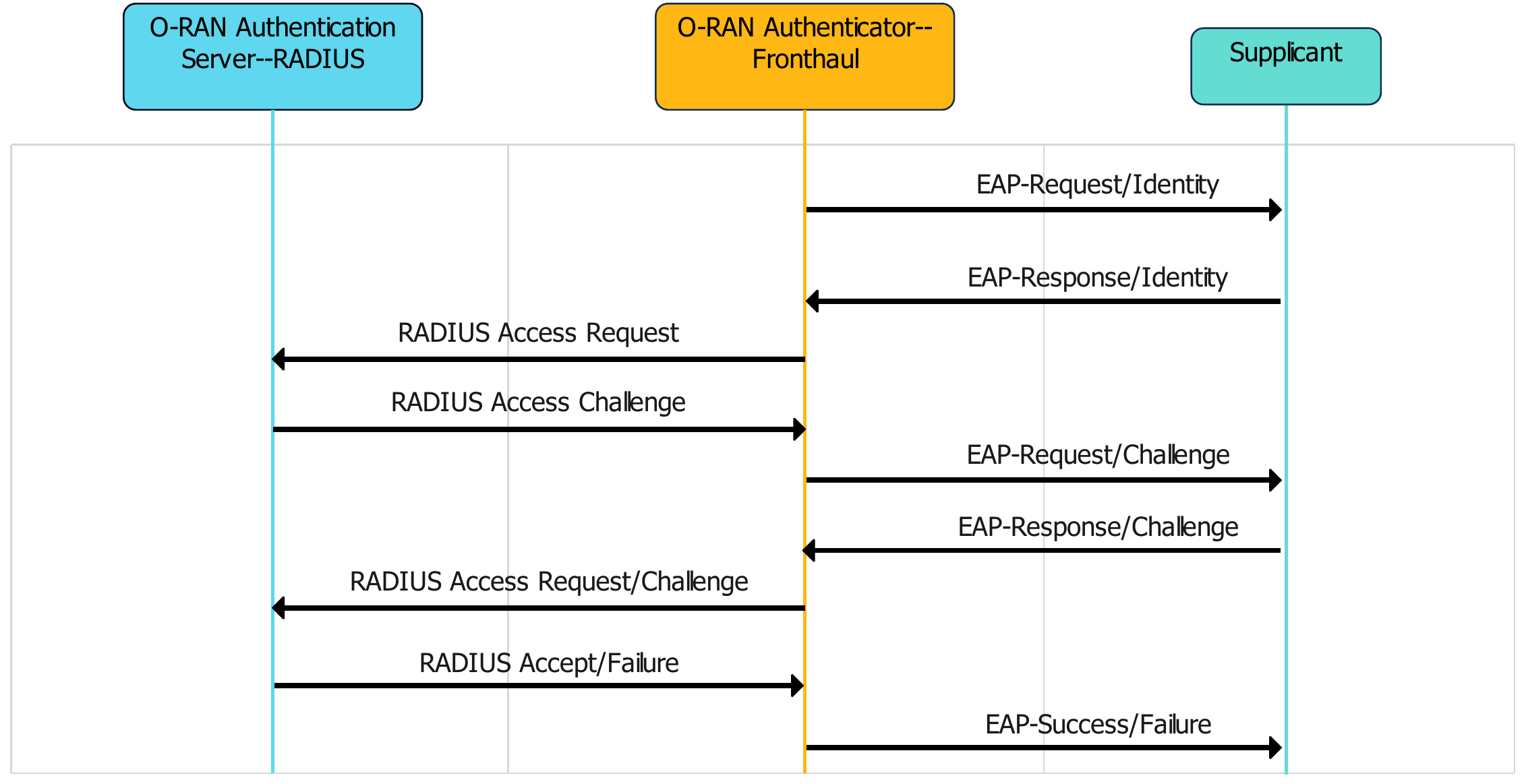}
    \label{fig:FigureORANIEEE802.1x}
\caption{The IEEE 802.1x authentication procedures \textcolor{black}{for secure 
MAC connectivity} (b).
}
\end{figure*}
\begin{itemize}
    \item \textbf{Certificate-Based Authentication:} 
    In order to authenticate each \textcolor{black}{network} component, a unique identifier and one or more credentials that need to be \textcolor{black}{securely stored} 
    are needed. A candidate approach 
    is 
    the mutual O-RAN component authentication based on \textcolor{black}{client/ server} certificates, such as X.509 certificates~\cite{rfc4210}. The X.509 is 
    \textcolor{black}{a} public key infrastructure (PKI) standard that is based on a strict hierarchical organization of the certificate authorities (CAs) in which the trust can only flow downwards. The CA acts as an effortlessly reachable trust anchor to all the components and interfaces within the O-RAN security architecture so that they can acquire the certificates during the secure socket layer/ 
    transport layer security (SSL/ 
    TLS) handshake. 
    After a 
    \textcolor{black}{X.509 certificate is issued and signed} by a trusted CA, the 
    user can be confident that the certificate owner or host name/ 
    domain has been validated. An additional benefit of this certificate-based authentication approach 
    is scalability. The PKI architecture is scalable in that it can secure huge amounts of exchanged messages by different components and interfaces \textcolor{black}{within} the O-RAN network and across the Internet. What enables this is that public keys can be distributed widely and openly without malicious actors being able to discover the private key required to decrypt the message.

    The X.509 certificate fields contain information about the identity that the certificate is issued to as well as the identity of the issuing CA. The standard fields include: version, serial number, algorithm information, issuer distinguished name, validity period of the certificate, subject distinguished name, and subject public key information. The O-RAN security specifications need to specify which identity fields in the X.509 \textcolor{black}{certificate} 
    should be checked during authentication, how these fields are formatted, and what the fields should be checked against to enable mutual authentication in an interoperable manner. The authenticated information is then used for authorization and policy control. O-RAN should make the profiling of the non-cryptographic identity fields in the X.509 \textcolor{black}{certificate} 
    stricter to enable interoperability among 
    vendors. 
    Fig.~4 presents an example of using mutual authentication based on X.509 certificates to authenticate the session between the PTP client and the authentication server in the O-RAN system.

   \vspace{+1 mm}
    \item \textbf{Port-Based Authentication:}
   The openness, multi-vendor support, 
   and the 7.2x split 
   \textcolor{black}{require that various clients access the northbound components and interfaces of O-RAN}. 
   Therefore, it is mandatory to enable harmonious authentication, authorization, and cryptographic key agreement mechanisms to support secure communication and network access in a point-to-point local area network (LAN) for the different client segments across the open fronthaul. 
   The IEEE 802.1x port-based access control mechanism is a candidate authentication scheme that can be implemented to rectify access by handling the transmission and reception of unidentified or unauthorized entities \textcolor{black}{over} the O-RAN fronthaul interface at the MAC layer and \textcolor{black}{thus avoid} consequent network disruption, such as 
   \textcolor{black}{service or data loss}.  
   
   IEEE 802.1x \textcolor{black}{is 
   a 
   framework} on top of which different authentication techniques, such as certificate-based one-time passwords and smartcard \textcolor{black}{readers}~\cite{IEEE802.1x} can be employed. 
   It \textcolor{black}{implements} 
   the extensible authentication protocol (EAP) 
   for transporting the authentication information and the verification of various authentication mechanisms 
   \textcolor{black}{through} challenge-response. 
   One of the main benefits \textcolor{black}{of IEEE 802.1x} 
   is that most of the \textcolor{black}{processing 
   happens} on the side of the clients that need access. 
   The main components of IEEE 802.1x are the supplicant, authentication server, and authenticator. \textcolor{black}{The} supplicant is the new user or client who wants to be authenticated, the authentication server is the server that validates the credentials transferred by the supplicant for determining its access, typically a RADIUS server, and the authenticator is the intermediate node that controls the communication between the supplicant and authentication server. 
   \textcolor{black}{Fig.~5 shows the corresponding message exchanges between the supplicant, O-RAN fronthaul authenticator, and the O-RAN authentication server for establishing secure MAC connectivity.} 

   \begin{figure}[ht]
    \centering
    \includegraphics[width=0.49\textwidth]{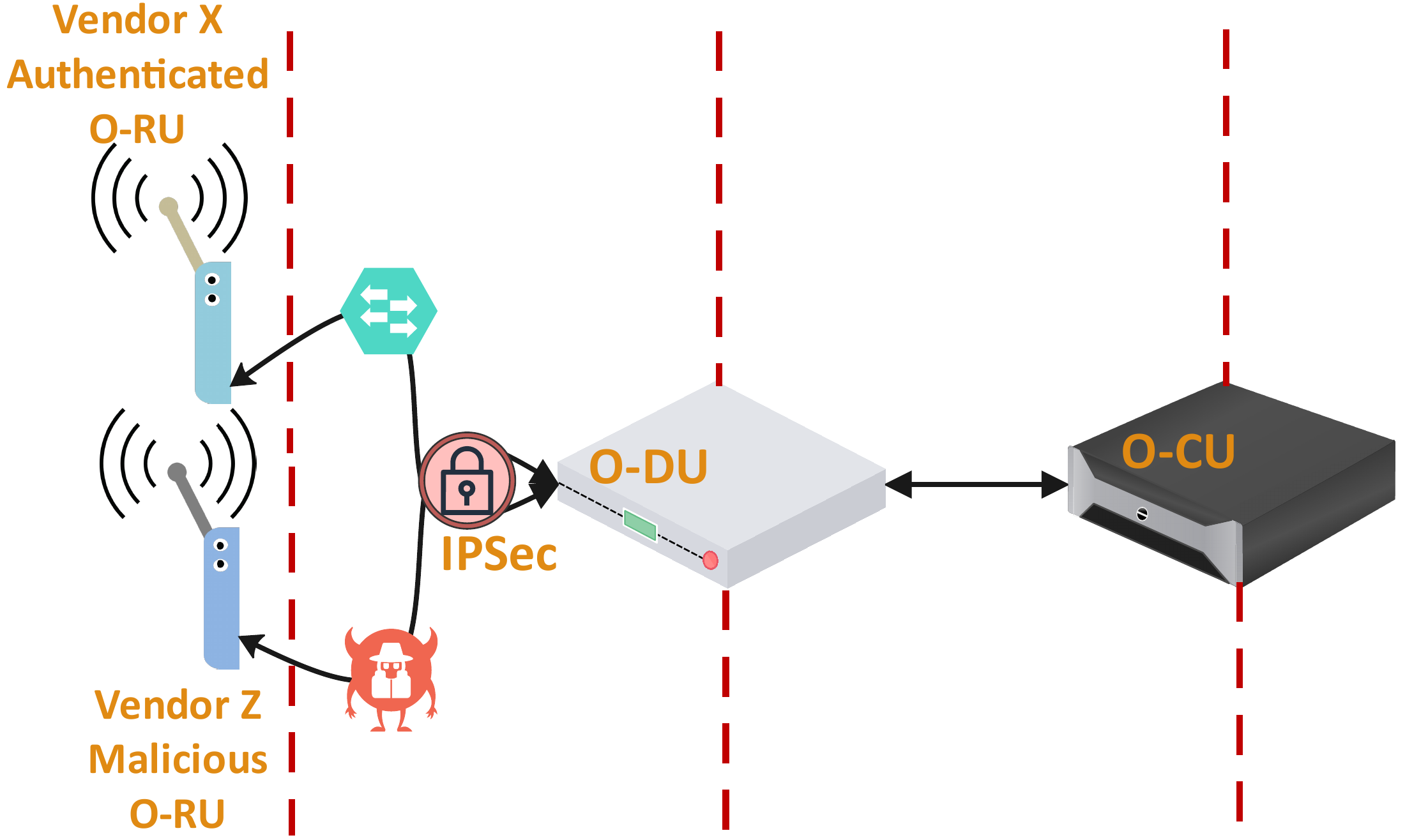}
    \label{fig:FigureORANIPsecSec}
    \caption{\textcolor{black}{Packet Tracer IPsec simulation scenario}}
\end{figure}
      \begin{figure}[h!]
    \centering
    \includegraphics[ width=0.49\textwidth]{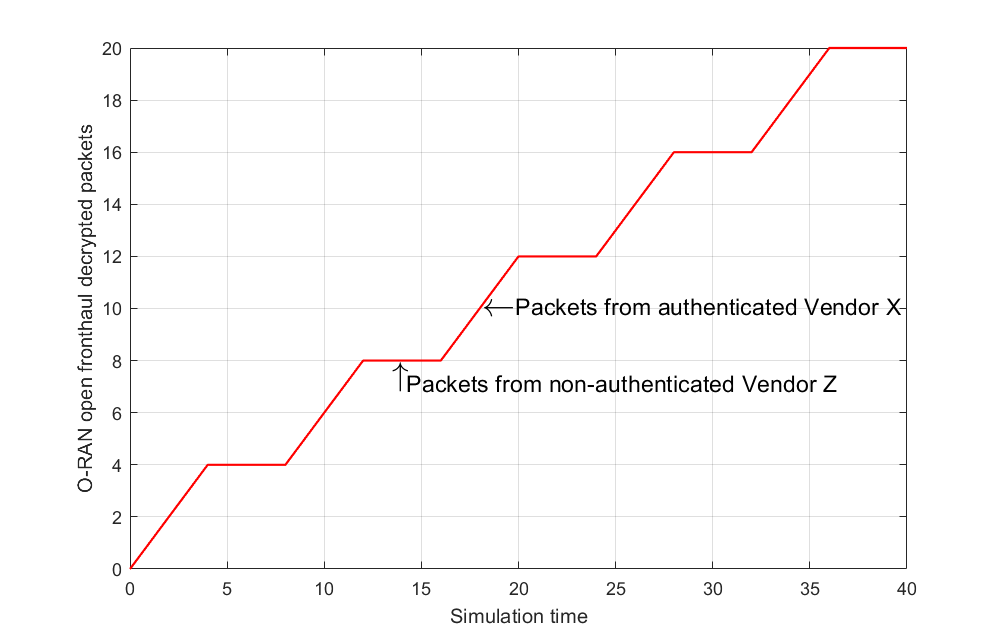}
    \label{fig:FigureORANPKTRACERESULT}
\caption{\textcolor{black}{The number of decrypted packets at the open fronthaul network termination for authenticated and malicious O-RU traffic over time.} 
}
\end{figure}
   \vspace{+1 mm}
    \item \textbf{IP Security-Based Authentication:}
     \textcolor{black}{O-RAN facilitates having different components,} 
     such as the O-RU and O-DU, supplied and managed by independent vendors that may implement different security mechanisms. \textcolor{black}{Such heterogeneous security measures 
     can introduce system vulnerabilities.} 
     However, it is not practical to maintain such a network within O-RAN taking into consideration the challenges of managing the complex infrastructure across multiple vendors over geographically dispersed areas. \textcolor{black}{Alternatively}, Internet Protocol security (IPsec) \textcolor{black}{can be 
     implemented} over the fronthaul interface of O-RAN at the IP layer for securing host-to-host IP packets and all other higher-layer protocols. IPsec \textcolor{black}{is} used for securing 
     non-3GPP access to the 5G core network~\cite{Sec5G}. IPsec is designed based on an authentication header (AH), encapsulating security payload (ESP), and Internet key exchange (IKE) transport-level protocols to provide authentication for the source and content of IP packets and, optionally, cipher the payload data. These three protocols empower IPsec to provide authentication; 
     specifically, proof of data source, data integrity, and 
     \textcolor{black}{replay} protection.

     \textcolor{black}{ In order to 
     \textcolor{black}{validate} 
     the
benefits of IPsec, we have set up an O-RAN emulation platform.  Fig.~6 shows the network topology that we consider. We assume that there are two O-RUs coming from different vendors, Vendor X and Vendor Z, where the O-RU from Vendor X is authenticated and the O-RU from Vendor Z is not 
in the authenticated O-RU list. Both the authenticated O-RU and the non-authenticated and potentially malicious O-RU aim to connect to the O-CU via an O-DU through the open fronthaul. We implement the IPsec tunnel that allows traffic coming from authenticated sources while blocking any traffic coming from unauthenticated entities. Without loss of generality, we assume time division multiple access for multiplexing the two O-RU transmissions. 
Each transmission phase is a simulation time unit that is characterized by four packet transmissions. 
Fig.~7 plots the decrypted packets by the O-RAN fronthaul termination at the O-DU over time. 
It 
shows that the implemented IPsec mechanism at the open fronthaul 
blocks packets coming from the non-authenticated/ malicious O-RU and lets packets through from the authenticated radio.}

\end{itemize}

\section{Conclusions}
\label{sec:conclusions}
The new RAN technologies and paradigm shift toward open, disaggregated, interoperable, multi-vendor, and AI empowered hierarchical networks 
expand the attack surface, introducing new security challenges and resultant risks. This article surveys the security vulnerabilities and threats targeting the O-RAN components with a focused study on the open fronthaul interface and the associated management, synchronization, control and data planes. 
\textcolor{black}{Based on the identified threats}, we discuss 
countermeasures 
to improve the security related to the open fronthaul in a multi vendor O-RU/O-DU context. 
This paper is expected to \textcolor{black}{spur research and encourage} industry to establish 
guidelines, standards, and best practices for safe
and secure implementations of O-RAN's open fronthaul interface.

\section*{Acknowledgement}
This work was supported in part by the National Science Foundation under grant number 2120442.

\balance

\bibliographystyle{IEEEtran}
\bibliography{Refs,vuk}
\section*{Biographies}
\small
\noindent
\textbf{Aly Sabri Abdalla} (asa298@msstate.edu)
is a PhD candidate in the Department of Electrical and Computer Engineering at Mississippi State University, Starkville, MS, USA. His research interests are on scheduling, congestion control and wireless security for vehicular ad-hoc and UAV networks.

\vspace{0.2cm}
\noindent
\textbf{Vuk Marojevic} (vuk.marojevic@msstate.edu) is an associate professor in electrical and computer engineering at Mississippi State University, Starkville, MS, USA. His research interests include resource management, vehicle-to-everything communications and wireless security with application to cellular communications, mission-critical networks, and unmanned aircraft systems.
\end{document}